\documentclass{article}

\usepackage[latin1]{inputenc}
\usepackage[italian,english]{babel}
\usepackage[dvips]{graphicx}

\begin{document}
\selectlanguage{english}


\title{Is it possible to detect gravitational waves\\ with atom interferometers?}

\author{
G. M. Tino $^{1}$, F. Vetrano $^{2}$}

\date{\today}
\maketitle

\begin{abstract}
We investigate the possibility to use atom interferometers to detect
gravitational waves. We discuss the interaction of gravitational waves with an
atom interferometer and analyze possible schemes.
\end{abstract}

\vspace{5mm}


\footnotetext[1]{Dipartimento di Fisica and LENS-Universit\`{a} di Firenze,
INFN-Sezione di Firenze, Polo Scientifico, 50019 Sesto Fiorentino, Italy -
E-mail: guglielmo.tino@fi.infn.it}

\footnotetext[2]{Istituto di Fisica-Universit\`{a} di Urbino, INFN-Sezione di
Firenze, 61029 Urbino, Italy - E-mail: vetrano@fis.uniurb.it}

\vspace{5mm} \textwidth = 27pc \textheight = 43pc

\section{Introduction}\label{par:Intro}
 The direct detection of Gravitational Waves (GWs) is one of the most exciting scientific
goals because it would improve our understanding of laws governing
the universe and provide new means to observe it. The most
sensitive detectors which are already operating, under
construction or being planned are based on optical interferometers
\cite{Saulson1994,reviewGW,reviewLISA}. In most cases, however,
the sensitivities are only marginally sufficient to detect the
expected signals, detectors have large sizes ranging from a few km
on the Earth (Virgo, LIGO) to millions of  km in space (LISA) and
the operating frequency ranges are limited. Therefore it is of
great interest to investigate alternative schemes that can lead to
a higher sensitivity, smaller sizes and extend the frequency range
of the detectors.

In recent years, matter-wave interferometry with neutral atoms has undergone an
impressive development due to the increasing ability to control the internal
and external atomic degrees of freedom using laser manipulation methods
\cite{Berman1997,Chu2001,Peters2001,Borde2002}. Atom Interferometers (AIs) are
already competing with state-of-art optical interferometers in terms of
sensitivity. This was demonstrated experimentally for gravity acceleration
\cite{Peters1999}, gravity gradients \cite{McGuirk2002}, inertial and rotation
effects \cite{Gustafson2000,Stedman1997}. Other experiments, planned or
presently in progress, to investigate properties of gravitational field by AI
concern accurate measurements of G \cite{Bertoldi2006,Fixler2007},  test of the
equivalence principle \cite{Dimopoulos2006}, detection of the Lense-Thirring
effect \cite{HYPER2000}, deviations from the 1/r$^{2}$ Newtonian law for small
distances \cite{Ferrari2006}.

In analogy to optical interferometers, in atom interferometers atomic wave
packets are split and recombined giving rise to an interference signal.
Different schemes can be used for splitting, reflecting and recombining the
atoms. In a particular class of interferometers, which is the one relevant in
this paper, the separation of the atoms is achieved by inducing a transition
between internal states of the atoms by an electromagnetic field. The spatial
separation in this case is induced by the momentum recoil and the internal and
external states of the atoms become entangled. Another approach is to use
material gratings. This raises however different problems, both conceptual and
technical, such as the realization and handling of the required nano-structures
and it will not be considered here.

In this paper, we discuss the possibility to use AI to detect gravitational
waves. The interaction between matter waves and gravitational waves was already
investigated in \cite{Linet1976,Stodolsky1979,Anandan1982,Borde1983}. Recently,
due to the experimental advances, the interest was revived
\cite{Aspen2004,Chiao2004,Misner2006,Foffa2006,Delva2006}. The aim of the
present paper is to analyze possible  schemes for interferometers using light
fields as atom optics components. Compared to \cite{Delva2006}, where only the
Linet-Tourrenc contribution is considered in the eikonal approximation, we take
into account all the contributions in phase difference inside the
interferometer.

The paper is organized as follows.    In Sect.~\ref{par:ABCDph} we recall the
ABCD formalism for matter waves and apply it to the calculation of phase shift
in atom interferometers in the specific case of weak gravitational field when
the Hamiltonian is at most quadratic in coordinates and conjugate momenta. A
detailed discussion about Einstein and Fermi coordinates is presented in
appendix. We apply the results in Sect.~\ref{par:phcalc} to derive the phase
shift signal for two specific atom interferometer configurations. Finally, in
Sect.~\ref{par:exp} we discuss possible experimental schemes and evaluate the
sensitivity for the detection of gravitational waves.

\section{The ABCD matrices for matter waves and phase-shift formula for
atom interferometers}\label{par:ABCDph} In this section we recall the ABCD
formalism for matter waves and apply it to calculation of phase shift formula.
The discussion is based on the relativistic Schroedinger-type equation for atom
waves and its analysis in \cite{Borde2004} and references therein.

\subsection{The ABCD matrices for matter waves}\label{par:ABCD}
In the following, we assume that the Hamiltonian relative to the motion of the
center of mass is a quadratic polynominal of momentum
and position operators, as in most cases of relevance in AI:%
\begin{eqnarray} \label{eq:quadH}
&&H=  \frac{1}{2M^*}\vec{p}\cdot \stackrel{\Rightarrow}{\beta}(t) \cdot \vec{p}
+ \frac{1}{2} \vec{p} \cdot \stackrel{\Rightarrow}{\alpha}(t) \cdot \vec{q}+
 \nonumber\\
&&- \frac{1}{2} \vec{q} \cdot \stackrel{\Rightarrow}{\delta}(t) \cdot \vec{p} -
\frac{M^*}{2}\vec{q}\cdot \stackrel{\Rightarrow}{\gamma}(t) \cdot \vec{q} +
\vec{f}(t) \cdot \vec{p} - M^* \vec{g}(t) \cdot \vec{q}
\end{eqnarray}%
 where $\alpha , \, \beta, \, \gamma, \, \delta$ are
suitable square matrices coefficients for the quadratic terms (with $\delta =
-\tilde{\alpha}$, where the tilde indicates the transposed matrix); $g$ is the
gravity vector field and $f$ is an external vector field. $M^*$ is the
relativistic mass ($M^* = M_0/\sqrt{1-v^2/c^2},$ where $M_0$ is the rest mass).
The evolution of the wave packets by this Hamiltonian, via the Ehrenfest
theorem, can be obtained through Hamilton's equations
\cite{Borde2004,Antoine2003}:
\begin{eqnarray}\label{eq:Heq}
\frac{d\chi (t)}{dt} = \left(\begin{array}{c}
  \frac{dH}{dp}\\
  -\frac{1}{M^*} \frac{dH}{dq}
\end{array}\right) = \Gamma (t) \chi (t) + \Phi (t)
\end{eqnarray}
where
\begin{eqnarray}\label{eq:chi}
\chi (t) = \left( \begin{array}{c}
  q\\
  p/M^*
\end{array}\right)
\hspace{1cm}
 \Phi (t) = \left( \begin{array}{c}
  f(t)\\
  g(t)
\end{array}\right)
\end{eqnarray}%
and%
\begin{eqnarray}\label{eq:Gamma}
\Gamma (t) = \left( \begin{array}{cc}
  \alpha (t) & \beta (t) \\
  \gamma (t) & \delta (t)
\end{array}\right)
\end{eqnarray}%
The integral of Hamilton's equations can be written through the ABCD matrices
as
\begin{eqnarray}\label{eq:chisol}
\chi (t) = \left( \begin{array}{cc}
  A (t, t_0) & B (t, t_0) \\
  C (t, t_0) & D (t, t_0)
\end{array}\right) \left[\chi (t_0) +
\left( \begin{array}{c}
  \xi (t,t_0) \\
  \psi (t,t_0)
\end{array}
\right) \right]
\end{eqnarray}
where
\begin{eqnarray}\label{eq:intpart}
\left( \begin{array}{c}
  \xi (t,t_0) \\
  \psi (t,t_0)
\end{array}
\right) =
 \int_{t_0}^{t}\mathcal{M}(t_0,t')\Phi (t')dt'
\end{eqnarray}%
and
\begin{eqnarray}\label{eq:ABCDsol}
\mathcal{M} (t,t_0) = \left( \begin{array}{cc}
  A (t, t_0) & B (t, t_0) \\
  C (t, t_0) & D (t, t_0)
\end{array}\right) =
\mathcal{T} \textrm{exp} \int_{t_0}^{t} \Gamma (t') dt'
\end{eqnarray}%
with $\mathcal{T}$ the time ordering operator. A perturbative expansion leads
to \cite{Antoine2003}:
\begin{eqnarray}\label{eq:Mpert}
\mathcal{M} (t,t_0) = 1+  \int_{t_0}^{t} \Gamma (t') dt' + \int_{t_0}^{t}  dt'
\int_{t_0}^{t'} \Gamma (t')\Gamma (t'')dt'' + ...
\end{eqnarray}%

The Eq.~(\ref{eq:ABCDsol}) can be used to find the ABCD matrices which
determine the evolution of the wave packets in presence of the GW, $h$. If we
consider the simple case in which the GW is the only (weak) field, in the two
coordinate systems discussed in Appendix A, for a single Fourier component $
\Omega$ we have up to the first
order in $h(\Omega)$:\\%

\textit{Fermi coordinates}
\begin{eqnarray}\label{eq:FCsubmatr}
   &&\stackrel{\Rightarrow}{\alpha} \,=\, \stackrel{\Rightarrow}{\delta} \, = \, 0 \nonumber\\
   &&\stackrel{\Rightarrow}{\beta} \,=\, \stackrel{\Rightarrow}{1}\\
   &&\stackrel{\Rightarrow}{\gamma} \,=\, \frac{\Omega^2}{2}
   \stackrel{\Rightarrow}{h}\nonumber
\end{eqnarray}

\begin{eqnarray}\label{eq:ABCDRBFC}
&&A(t_2,t_1)=1- \gamma(\Omega)e^{i\Omega t_1} \left[\frac{e^{i\Omega
(t_2-t_1)}-1}{\Omega^{2}}
+\frac{t_2-t_1}{i\Omega}\right]  \nonumber\\
&&B(t_2,t_1)=(t_2-t_1)+\nonumber\\
&&+ \frac{\gamma(\Omega)}{\Omega^{2}}e^{i\Omega
t_1}\left[-(t_2-t_1)\left(e^{i\Omega(t_2-t_1)}+1\right)
+\frac{2\left(e^{i\Omega(t_2-t_1)}-1\right)}{i\Omega}\right]  \nonumber\\
&&C(t_2,t_1)= \gamma(\Omega)e^{i\Omega t_1} \left[\frac{e^{i\Omega
(t_2-t_1)}-1}{i\Omega}\right]
  \\
&&D(t_2,t_1)=1+ \gamma(\Omega)e^{i\Omega t_1} \left[\frac{(t_2-t_1)e^{i\Omega
(t_2-t_1)}}{i\Omega} +\frac{e^{i\Omega(t_2-t_1)}-1}{\Omega^{2}}\right]
\nonumber
\end{eqnarray}%
where $\gamma(\Omega)=h(\Omega)\Omega^{2}/2, h(\Omega)=\int h(t)\exp(-i\Omega
t)dt, i=\sqrt{-1}$ .\\

\textit{Einstein coordinates}
\begin{eqnarray}\label{eq:ECsubmatr}
   &&\stackrel{\Rightarrow}{\alpha}\, =\, \stackrel{\Rightarrow}{\delta} \,=\,
   \stackrel{\Rightarrow}{\gamma} \,= \,0 \nonumber\\
   &&\stackrel{\Rightarrow}{\beta} \,= \,\stackrel{\Rightarrow}{h}(t) - \stackrel{\Rightarrow}{\eta}
\end{eqnarray}

\begin{eqnarray}\label{eq:ABCDRBEC}
&&A(t_2,t_1)=1\nonumber\\
&&B(t_2,t_1)=(t_2-t_1)+ \frac{h(\Omega)}{i\Omega}e^{i\Omega
t_1}\left[e^{i\Omega(t_2-t_1)}-1\right]  \nonumber\\
&&C(t_2,t_1)=0 \\
&&D(t_2,t_1)=1 \nonumber
\end{eqnarray}

where $\stackrel{\Rightarrow}{\eta}$ is the Minkowskian matrix.

\subsection{The phase shift formula for atom interferometers}\label{par:ph}
The total phase difference between the two arms, $s$ and $i$, of an atom
interferometer can be expressed as the sum of three terms: the difference in
the action integral along each path; the difference in the phases imprinted by
the beam-splitters on the atom waves; the contribution from the splitting of
the wave packets at the exit of the interferometer \cite{Borde2004}:
\begin{eqnarray}\label{eq:Del1}
\Delta \varphi &=& \frac{1}{\hbar}\sum_{j=1}^{N} [S_{s}(t_{j+1},t_j) -
S_{i}(t_{j+1},t_j)] + \nonumber\\
&&+\sum_{j=1}^{N}[(k_{sj} q_{sj} - k_{ij} q_{ij}) - (\omega_{sj} -
\omega_{ij})t_j
+ (\vartheta_{sj}- \vartheta_{ij})]+ \nonumber\\
&&+ \frac{1}{\hbar}[p_{sD}(q-q_{sD})-p_{iD}(q-q_{iD})]
\end{eqnarray}
where $S_{sj}=S_s (t_{j+1},t_j)$ and $S_{ij}=S_i (t_{j+1},t_j)$ are the action
integrals along $s$ $(i)$ path;  $k_{sj}(k_{ij})$ is the momentum transferred
to the atoms by the j-th beam-splitter along the $s$ $(i)$ arm; $q_{sj}$ and
$q_{ij}$ are the coordinates of the beam splitter/atom interaction;
$\omega_{sj} (\omega_{ij})$ is the angular frequency of the laser beam;
$\vartheta_{sj} (\vartheta_{ij})$ is the phase of the laser beam at the j-th
interaction with the atom; $D$ is the exit port.

Assuming the same input point for the two arms and using the "mid point"
property \cite{Borde2002} in integrating over the space at the output, the
phase shift difference $\Delta \varphi$ between the two arms (s,i) for an
interferometer with N beam splitters can be written as:
\begin{eqnarray}\label{eq:Del}
\Delta \varphi &=& \sum_{j=1}^{N} (k_{sj} - k_{ij})\frac{q_{sj} + q_{ij}}{2}+
\sum_{j=1}^{N} (\omega_{sj} -
\omega_{ij})t_j + \nonumber\\
&&+ \sum_{j=1}^{N} (\vartheta_{sj} - \vartheta_{ij}) +
\sum_{j=1}^{N}\frac{(M_{sj} - M_{ij}) c^2}{\hbar}\tau_j
\end{eqnarray}
where $M_{sj}(M_{ij})$ is the mass of the atom in the $s$ $(i)$ arm and
$\tau_j$ is a proper time at the j-th interaction.

\section{Phase calculation for different AI geometries}\label{par:phcalc}
In this section, we use the ABCD formalism to find the resulting phase shift
$\Delta \varphi$ for typical atom interferometer schemes \cite{Berman1997}. The
approach is in the frequency space of complex Fourier transform in order to
describe both the amplitude and the phase of the resulting $\Delta \varphi$. We
consider here the trapezoidal interferometer, first suggested in
\cite{Borde1989}, and the parallelogram-shaped interferometer, both in Fermi
and Einstein coordinates, retaining only terms up to the first order in $h$.

\subsection{Trapezoidal  AI}\label{par:RB}
\begin{figure}[ht]
\centering
\includegraphics[width=0.8\textwidth]{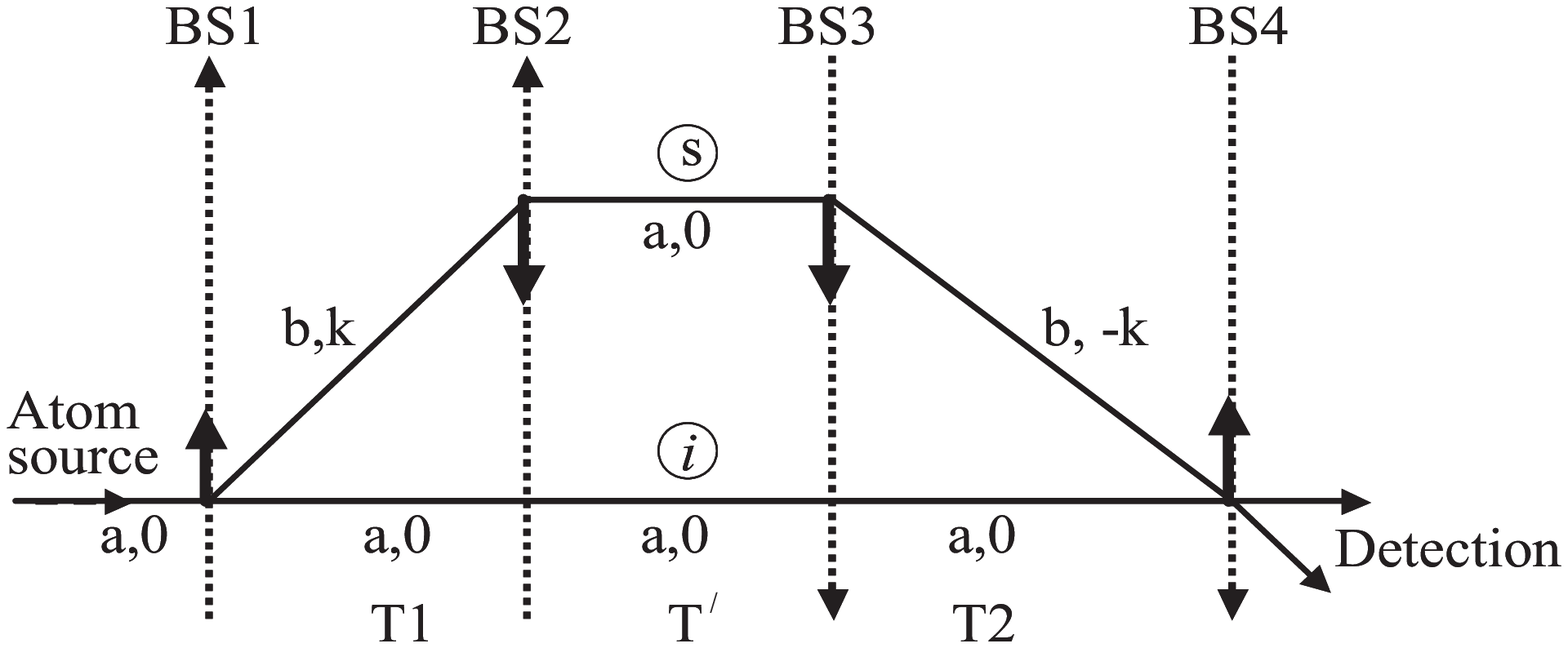}
\caption{Scheme of the trapezoidal interferometer. Dotted arrows represent
laser beams acting as beam-splitters (BS1-BS4); bold continuous arrows show the
relevant momentum transferred to the atom; a: ground internal atomic state; b:
excited internal atomic state; \textbf{k}: transferred momentum (in $\hbar$
units). In the text $T_1$ = $T_2$ = $T$ and $T'$ $\rightarrow $ 0
.}\label{fig:RB}
\end{figure}

The scheme of a trapezoidal interferometer is shown in Fig.~\ref{fig:RB}. The
interaction of the atom with two counter-propagating pairs of copropagating
beams (BS1-BS4) gives rise to a trapezoidal-shaped closed circuit. By using
Eq.~(\ref{eq:Del}), we obtain the expression of the phase shift for the two
arms
(s,i) of the interferometer:%
\begin{eqnarray}\label{eq:DelRB}
&&\Delta \varphi= k_1 q_1 + \frac{1}{2} k_{s2} (q_{s2,b} +
q_{i2,a})+ \frac{1}{2} k_{s3} (q_{i3,a} +
q_{s3,a}) + \frac{1}{2} k_{4} (q_{s4,b} + q_{i4,a}) \nonumber\\
&&+ 2(\Omega_L T - \Omega_{ba} \tau) + \vartheta _1 -\vartheta _2 + \vartheta
_3 -\vartheta _4
\end{eqnarray}%
where $\Omega_L$ is the laser frequency and $\Omega_{ba}$ is the
frequency of the atomic transition involving ground (a) and
excited (b) states; $\vartheta _i$ are the proper laser phases.
Expressing all the $q_j$ coordinates through the ABCD matrices,
the phase difference $\Delta \varphi$ at the output of the interferometer is then given by:%
\begin{eqnarray}\label{eq:DRBFC}
&&\Delta \varphi = k_1q_1[1-2A(T,0)+A(2T,0)]+\nonumber\\
&&+\frac{k_1}{2}[B(2T,0)-2B(T,0)]\left(\frac{p_1}{M_b}+\frac{p_1}{M_a} +\frac{\hbar k_1}{M_b}\right)\\
&&-k_1B(2T,T)\frac{\hbar k_1}{M_b}+2\Omega_L T -2\Omega_{ba} \tau + \vartheta
_1 -\vartheta _2 + \vartheta _3 -\vartheta _4 \nonumber
\end{eqnarray}%
where M$_{a}$ and M$_{b}$ are the masses of the atom in the ground and excited
state, respectively; the expression for the A, B, C, D matrices depends on the
coordinate system.

Let's first consider Fermi coordinates. From Eq.~(\ref{eq:DRBFC}) and
Eq.~(\ref{eq:ABCDRBFC}), we obtain for the phase difference for this
configuration
\begin{eqnarray}\label{eq:DRBFCOm}
&&\Delta \varphi(\Omega)= -\frac{\Omega \, h(\Omega)}{2}
\left(\frac{1}{M_a}+\frac{1}{M_b}\right)T^{2}k_1p_1 \cdot\nonumber\\
&& \cdot \left\{ \left[\sin\Omega T\left(\frac{\sin(\Omega T/2)}{\Omega T/2}
\right)^{2}+\frac{\cos 2\Omega T - \cos \Omega T}{\Omega T}
 \right] +  \right. \nonumber\\
&&\left. + \, i\left[\frac{\sin 2\Omega T - \sin \Omega T}{\Omega T} -
\cos\Omega T\left(\frac{\sin(\Omega T/2)}{\Omega
T/2} \right)^{2}\right]\right\}+\nonumber\\
&&+  h(\Omega) \frac{\hbar k_1^{2}T}{M_b} \left[\left(1-\frac{\sin\Omega T
}{\Omega T}\right)\cos\Omega T +
i\left(1-\frac{\sin\Omega T}{\Omega T}\right)\sin\Omega T \right]+ \nonumber\\
&& +\frac{\Omega^{2}h(\Omega)}{2} T^{2}k_1q_1
\left[\frac{\sin(\Omega T/2)}{\Omega
T/2} \right]^{2}\left(\cos\Omega T + i\sin \Omega T \right) + \\
&&+ 2 \left(\Omega_L - \frac{\hbar k_1^{2}}{2 M_b}\right)T - 2\Omega_{ba} \tau
+ \vartheta _{1F} -\vartheta _{2F} + \vartheta _{3F} -\vartheta _{4F} \nonumber
\end{eqnarray}
Considering Einstein coordinates, from Eq.~(\ref{eq:DRBFC})
and Eq.~(\ref{eq:ABCDRBEC}), we obtain
\begin{eqnarray}\label{eq:DRBECOm}
&&\Delta \varphi(\Omega) = -\frac{\Omega h(\Omega)}{2}
\left(\frac{1}{M_a}+\frac{1}{M_b}\right)T^{2}k_1p_1 \cdot \nonumber\\
&& \cdot \left[\sin\Omega T\left(\frac{\sin(\Omega T/2)}{\Omega
T/2} \right)^{2}- i\cos\Omega
T\left(\frac{\sin(\Omega T/2)}{\Omega T/2} \right)^{2} \right] + \\
&& -\frac{\hbar k_1^{2}T}{M_b}h(\Omega)\left(\cos\Omega T \frac{\sin \Omega
T}{\Omega T} + i \sin \Omega T
\frac{\sin \Omega T}{\Omega T}\right) + \nonumber\\
&& + 2 \left(\Omega_L - \frac{\hbar k_1^{2}}{2 M_b}\right)T - 2\Omega_{ba} \tau
+ \vartheta _{1E} -\vartheta _{2E} + \vartheta _{3E} -\vartheta _{4E} \nonumber
\end{eqnarray}
where $\vartheta _{iE}$ are the laser phases in the Einstein coordinates
system. They are different because in Einstein coordinates the index of
refraction for the vacuum
 is varying with
$h(t)$ or, which is the same, we have an extra (Fourier transformed)
contribution $\delta k \cong k[h(t,\Omega)/2]$exp$(i \Omega t)$ to the momentum
transfered at the beam-splitter positions as a consequence of the apparent
photon velocity $v \cong c [1 + (h/2)] $. This can be accounted for in the
phase terms leading to an extra term $-\delta k (t,\Omega)$ in
Eq.~(\ref{eq:DelRB}), as in spatial beam-splitters \cite{Borde2004}. By
inserting these laser phases and using the coordinates transformation rules in
GR between the two systems here considered (see Appendix B), the resulting
phase shift difference is coincident with the one obtained in Fermi
coordinates, as expected for a scalar quantity which is a physical result in
spite of different descriptions.

\subsection{Parallelogram-shaped AI}\label{par:MZ}
\begin{figure}[ht]
\centering
\includegraphics[width=0.8\textwidth]{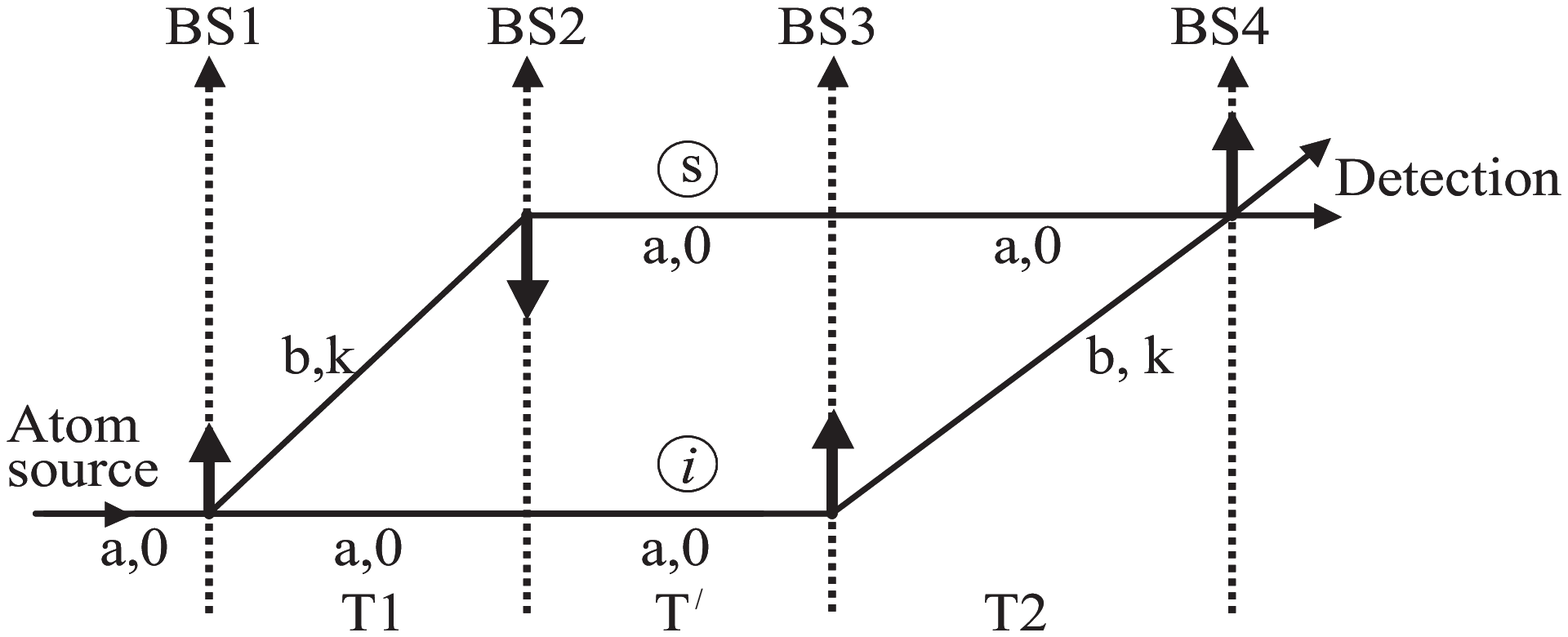}
\caption{Scheme of a parallelogram interferometer. Dotted arrows represent
laser beams acting as beam-splitters (BS1-BS4); bold continuous arrows show the
relevant momentum transferred to the atom; a: ground internal atomic state; b:
excited internal atomic state; \textbf{k}: transferred momentum (in $\hbar$
units). In the text $T_1$ = $T_2$ = $T$ and $T'$ $\rightarrow $ 0 .}
\label{fig:Diam}
\end{figure}

The scheme of a parallelogram-shaped AI is shown in Fig.~\ref{fig:Diam}. In
this case, the interaction of the atom with four copropagating laser beams
gives rise to a parallelogram-shaped closed circuit.

The phase difference at the output of this interferometer is given by:%
\begin{eqnarray}\label{eq:DDIFC}
&&\Delta \varphi = k_1q_1[1-2A(T,0)+A(2T,0)]+\nonumber\\
&&+\frac{k_1}{2}[B(2T,0)-2B(T,0)]\left( \frac{p_1}{M_a}+\frac{p_1}{M_b} +\frac{\hbar k_1}{M_b}\right) +\\
&&+ \frac{\hbar k_1^{2}}{2M_b}B(2T,T)\left[D(T,0)-1\right]\epsilon + \vartheta
_1 -\vartheta _2 - \vartheta _3 +\vartheta _4 \nonumber
\end{eqnarray}
up to the first order in $\epsilon$, where $\epsilon= (M_b-M_a)/M_a$. The only
difference from the case of a trapezoidal interferometer (Eq.~(\ref{eq:DRBFC}))
is in the recoil term, which is proportional to the relative energy difference
between ground and excited states. Using Fermi coordinates, from
Eq.~(\ref{eq:ABCDRBFC}) we
obtain:%
\begin{eqnarray}\label{eq:DDIFCOm}
&&\Delta \varphi(\Omega)= -\frac{\Omega \, h(\Omega)}{2} T^{2}k_1 \left(
\frac{p_1}{M_a}+\frac{p_1}{M_b}+\frac{\hbar k_1}{M_b} \right)
\cdot\nonumber\\
&& \cdot \left\{\left[\sin\Omega T\left(\frac{\sin(\Omega T/2)}{\Omega T/2}
\right)^{2}+\frac{\cos 2\Omega T - \cos \Omega T}{\Omega T}
 \right]+ \right. \nonumber\\
&& \left. + i\left[\frac{\sin 2\Omega T - \sin \Omega T}{\Omega T}
- \cos\Omega T\left(\frac{\sin(\Omega T/2)}{\Omega
T/2} \right)^{2}\right]\right\}+ \nonumber\\
&&+\frac{\Omega^{2}h(\Omega)}{2}
T^{2}k_1q_1\left(\frac{\sin(\Omega T/2)}{\Omega T/2}
\right)^{2}\left(\cos\Omega T + i\sin \Omega T \right) + \\
&&+ \frac{\epsilon}{2}\frac{\hbar
k_1^{2}}{2M_b}h(\Omega)T^{2}\left[\left(\sin\Omega T -\frac{1- \cos\Omega
T}{\Omega T} \right)+ i\left(\frac{\sin \Omega T }{\Omega T} - \cos \Omega T\right)\right] + \nonumber\\
&& +\vartheta _{1F} -\vartheta _{2F} - \vartheta _{3F} +\vartheta _{4F}
\nonumber
\end{eqnarray}

Considering Einstein coordinates, we obtain:%
\begin{eqnarray}\label{eq:DDIECOm}
&&\Delta \varphi(\Omega))= -\frac{\Omega h(\Omega)}{2}
T^{2}k_1\left(\frac{p_1}{M_a}+\frac{p_1}{M_b}+\frac{\hbar k_1}{M_b}\right)
\cdot
\\
&& \cdot \left[\sin\Omega T\left(\frac{\sin(\Omega T/2)}{\Omega
T/2} \right)^{2}- i\cos\Omega
T\left(\frac{\sin(\Omega T/2)}{\Omega T/2} \right)^{2} \right] + \nonumber\\
&& + \vartheta _{1E} - \vartheta _{2E} - \vartheta _{3E}
+\vartheta _{4E} \nonumber
\end{eqnarray}

The same considerations of paragraph \ref{par:RB} apply in this case about the
identity of the results in the two descriptions.

\section{Possible schemes and expected sensitivities}\label{par:exp}

The results in previous section provide the phase shift at the
output of the atom interferometer induced by a gravitational wave
with amplitude $h$ and frequency $\Omega$ for the typical schemes
considered. In order to determine the sensitivity of the
interferometer, that is the minimum detectable amplitude
$h(\Omega)$, we assume shot-noise-limited detection of the atoms,
corresponding to a phase noise given by $\Delta
\varphi(\Omega))=\eta/\sqrt{\dot{N}}$ where $\eta$ is a detection
efficiency and $\dot{N}$ is the atoms flow at the detector. The
resulting sensitivity $h(\Omega)$ (at S/N ratio equal 1)
can be written as%

\begin{equation}\label{eq:gensens}
  h(\Omega)= \frac{ \eta }{\sqrt{\dot{N}}} \,\,
  \frac{1}{   f(\Omega) \Sigma }
\end{equation}%
where $\Sigma$ is a scale factor and $f(\Omega)$ is the resonance function of
the interferometer. Neglecting the clock and recoil terms, from
Eq.~(\ref{eq:DRBFCOm}) we obtain:
\begin{equation}\label{eq:homega}
  h(\Omega)= \frac{\eta\hbar}{p_{T}L \sqrt{\dot{N}}} \,\, \frac{1}{f(\Omega)}
\end{equation}%
where $L$ is the characteristic linear dimension of the interferometer,  $p_T =
mv_T$ with $v_T$ the transversal velocity acquired by the atoms in the
splitting process and $2/m =
1/M_a + 1/M_b$, and where%
\begin{eqnarray}\label{eq:fomega}
  f(\Omega)= \Omega T \left\{ \left[ \sin (\Omega T)
\left( \frac{\sin (\Omega T)/2}{\Omega T/2}\right)^2 + \frac{\cos
(2 \Omega T) - \cos ( \Omega T)}{\Omega T}
\right] + \right. \nonumber \\
 \left. + \, i  \left[ \frac{\sin (2 \Omega T) - \sin ( \Omega T)}{\Omega T} - \cos (\Omega T) \left(
\frac{\sin (\Omega T)/2}{\Omega T/2}\right)^2 \right]
    \right\}
\end{eqnarray}%
 From Eq.~(\ref{eq:homega}), it is evident
that in order to achieve the required sensitivity while keeping a sufficiently
large detection bandwidth it is necessary to realize large values of $L$ and
$p_{T}$.

In order to evaluate the performance of these new detectors for GWs, we
analyzed a few specific cases. It is important to notice that in this analysis
we did not treat other noise sources that, as in optical GW detectors, can
affect the performance of the AI detector. Examples are the suspension of
optics required for the manipulation of the laser beams or the phase noise of
the laser itself. Based on the work for optical GW detectors and progress in
ultrastable lasers for future optical clocks, suitable laser sources and
suspension systems can be envisaged. A detailed analysis of the overall noise
budget, including technological aspects, is beyond the scope of the present
paper.

\begin{figure}[ht]
\centering
\includegraphics[width=\textwidth]{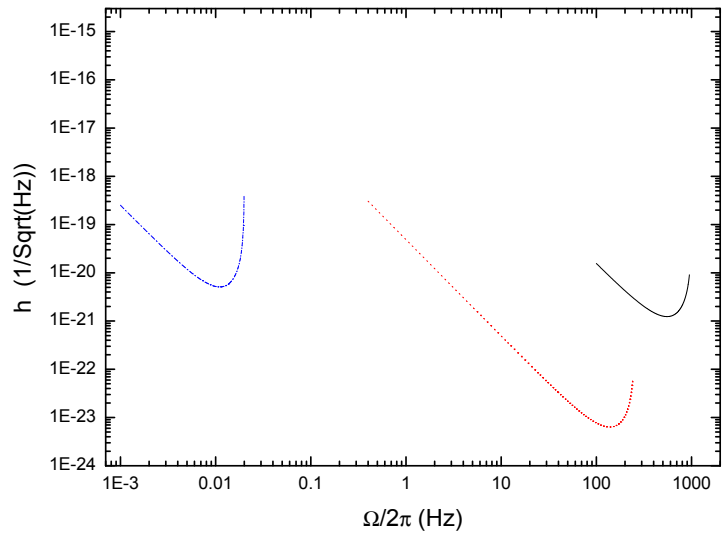}
\caption{Sensitivity curves of GW detectors based on atom interferometry for
the three parameter sets discussed in the text: $L = 10^3$ m, $v_L = 10^6$ m/s
(continuous line);  $L = 2 \cdot 10^5$ m, $v_L = 5 \cdot 10^7$ m/s (dotted
line); $L = 50$ m, $v_{L} = 1$ m/s (dashed-dotted line).}
\label{fig:Sensitivity}
\end{figure}

Let us consider first an atom interferometer based on a fast beam of hydrogen
atoms.
If we take $T = 10^{-3}$~s and a length $L = 10^{3}$~m, similar to present
optical interferometers detectors,
we have $v_{L} = 10^{6} $ m/s.  As shown in Fig.\ref{fig:Sensitivity}, in this
case a sensitivity $h(\Omega)$ of about $10^{-21}/Hz^{1/2}$ is achieved for
$v_{T} \approx 10$ m/s with a flux of $10^{18}$ atoms/s in the atomic beam
\cite{Scoles1988}. The recoil velocity for a hydrogen atom absorbing a
$\textrm{Ly}-\alpha$ photon is $v_{rec} = 3.3$ m/s. Although the absorption of
a photon followed by spontaneous emission destroys coherence and cannot be used
to deflect atomic trajectories in an interferometer, it is conceivable to use
two-photon Raman transitions between the two hyperfine levels of H ground
state.  A single Raman pulse transfers a velocity $v_{T} = 2 v_{rec}$. Raman
transitions have already been used in AI based on alkali atoms
\cite{Berman1997} and the possibility to use multiple Raman pulses sequences to
increase the enclosed area and the resulting sensitivity was also demonstrated
\cite{McGuirk2000}. A practical limitation at present would be the required
power $(\approx 10~W)$ of laser radiation at the Ly$\alpha$ wavelength. This is
orders of magnitude larger than what can be presently achieved as cw radiation
\cite{Eikema1999} but closer to what is produced in pulsed mode
\cite{Setija1993}. An alternative scheme is the excitation of a two-photon
transition from the ground state to long lived excited states
\cite{Gross1998,Hijmans2000}. A large recoil can be transfered by combining
$1s-2s$ excitation with optical transitions from the $2s$ state to high lying
$p$ states \cite{Heupel2002}. Such a scheme is also compatible with a ground
based apparatus because of the negligible vertical displacement of the atomic
beam during the short total time of flight. Other cases we considered for fast
beams of heavier atoms do not meet the requirements for the scheme we
considered because of the difficulty to transfer a large enough transverse
momentum to the atoms while keeping  T small enough in order to keep a large
bandwidth. An improvement in sensitivity, at the expense of a reduced
bandwidth, could be achieved by increasing T and correspondingly the linear
dimensions of the interferometer. In this case, however, a gravity-free
apparatus in space should be considered (Fig.\ref{fig:Sensitivity}).

A different case we considered is an interferometer based on cold atoms. In
this case, it is more useful to rewrite Eq.~(\ref{eq:homega}) in the form:
\begin{equation}\label{eq:homegaslow}
  h(\Omega)= \frac{\eta\hbar}{\sqrt{\dot{N}}}
  \left(\frac{p_{L}}{p_{T}}\right)\frac{T}{m L^{2} }\frac{1}{f(\Omega)}
\end{equation}
where $m$ is the atomic mass. It is apparent that in this scheme, by relaxing
the constraint on $T$, the sensitivity is better the larger is the value of $m$
and the smaller is $p_{L}$. As an example, if $v_{L} = 1$ m/s = $2v_T$ , $L =
50$ m and $m \approx 10^{2}$ a.m.u., a sensitivity of about $10^{-21}/Hz^{1/2}$
results at  frequencies around 10 mHz (Fig. \ref{fig:Sensitivity}). The long
time of flight $T \approx 50$ s, for which a gravity-free scheme would be
required, leads of course to a narrower bandwidth.


\section{Conclusions}\label{par:concl}
We investigated the possibility of detecting gravitational waves using atom
interferometers based on light fields as beam-splitters. The phase shift at the
output of the interferometers was calculated for presently known schemes using
both Einstein and Fermi coordinates. Considering sensitivities of the same
order of magnitude as the ones of present optical gravitational wave detectors,
we estimated the resulting values for relevant parameters. The results show
that dedicated technological developments would be needed to achieve the
required values which are beyond those presently available. New schemes for
atom interferometers, beam splitters, and high flux coherent atomic sources can
lead to an increase in sensitivity and make atom interferometers
competitive with other gravitational wave detectors.\\

\section*{Acknowledgments}
The authors acknowledge useful discussions with Ch. J. Bord\'{e}.

\section*{Appendix A: Einstein vs Fermi coordinates}\label{par:EvsF}


In general relativity (GR) all the coordinates systems are $\textit{a priori}$
equivalent. The predicted physical results do not depend on the specific
coordinates system although different descriptions depend
on coordinates systems. Generally speaking, a change of coordinates is defined by any set of functions%
\begin{equation}\label{eq:coord}
x^{\alpha}=x^{\alpha}(y^\beta)\longleftrightarrow y^{\alpha}=y^{\alpha}(x^
\beta) \;\;\; \alpha,\beta=0,1,2,3
\end{equation}%
where the invertibility is guaranteed if and only if%
\begin{equation}\label{eq:invert}
 \textrm{det} \left( \frac{\partial x^{\alpha}}{\partial y^{\beta}} \right) \neq 0
\end{equation}%
In the following, we refer to the case in which deviations from the Minkowsky
space of special relativity are due only to GWs in the weak field
approximation, that is%
\begin{equation}\label{eq:h}
g_{\mu\nu}(x) = \eta_{\mu\nu}+h_{\mu\nu}(x)
\end{equation}
with $|h_{\mu\nu}|<<1$.
We assume that no other field is present. In this case, the linearized Einstein
field equations admit a plane wave solution for  $h_{\mu\nu}$. It is always
possible to choose Gaussian synchronous
coordinates \cite{Stephani} in which%
\begin{equation}\label{eq:g}
g_{0\mu} = g_{\mu 0} = (1,0,0,0)
\end{equation}
Let's choose the particular coordinates system in which two particles A and B,
at rest in Minkowsky system, remain at rest even in presence of a GW. These are
called Einstein coordinates
(EC) and the related metrics can be written as%
\begin{equation}\label{eq:metric}
ds^{2}= c^{2}dt^{2} - (1-h_{+}) dx^{2} - (1+h_{+})dy^{2} - dz^{2} +
2h_{\times}dxdy
\end{equation}%
for a wave propagating along the z axis; $h_{+}=h_{+}(ct-z)$ and
$h_{\times}=h_{\times}(ct-z)$ are the amplitudes of the two polarizations
states. The world-lines of a free particle are geodesics \cite{Misner}. It is
important to note that the proper distance between two particles A and B always
at
rest in this system is varying with the amplitude of the GW \cite{Brillet1985}:%
\begin{equation}\label{eq:d2}
d^{2} = d_{0}^{2}+h_{ij}(t)(x_{B}^{i}-x_{A}^{i})(x_{B}^{j}-x_{A}^{j})
\end{equation}%
where $ d_{0}^{2} = (x_{B}^{i}-x_{A}^{i})(x_{B}^{i}-x_{A}^{i})$ in the
hypothesis that the particles A and B are close enough to consider $h_{ij}$
depending only on t. From another point of view, we can say that the flight
time of a photon from A to B and back is varying, or that we have an index of
refraction of the vacuum which is varying with the perturbation $h_{ij}(t)$.

The Einstein coordinates are formally the most convenient to describe plane
GWs; they can be considered as a "wave system (TT gauge)". This is not an
intuitive system, however, for measurements in a laboratory;  as an extension
of "classical" approach, indeed, we search for inertial systems in which it is
possible to preserve the Newtonian idea of "rigid stick" and related
measurement method. Fermi coordinates (FC) are the best approximation to such a
"Galilean system" \cite{Misner,Manasse1963}. We choose a "fiducial observer"
(free falling observer) at rest in the origin; the system is related to the
geodesic line $x=y=z=0$. This is a Minkowskian system if we disregard small
terms as $h_{+}$ and $h_{\times}$ \cite{Grishchuk}. The spatial axis are built
as locally orthogonal coordinate lines whose direction can be checked by
gyroscopes \cite{Manasse1963,Brillet1985,Grishchuk}. The transformation
between the two systems EC and FC are \cite{Brillet1985,Grishchuk}:%
\begin{eqnarray}\label{eq:ECFC}
&&X=x - \frac{1}{2}h_{+}x+\frac{1}{2}h_{\times}y \nonumber \\
&&Y=y + \frac{1}{2}h_{+}y+ \frac{1}{2}h_{\times}x \\
&&Z=z + \frac{1}{4c}\dot{h}_{+}(x^{2}-y^{2})+\frac{1}{2c}\dot{h}_{\times}xy \nonumber \\
&& T=t +\frac{1}{4c^2}\dot{h}_{+}(x^{2}-y^{2})+\frac{1}{2c^2}\dot{h}_{\times}xy
\nonumber
\end{eqnarray}
where X, Y, Z, T are Fermi coordinates, x, y, z, t are Einstein coordinates,
and the dot indicates the time derivative.

Two particles $A$ and $B$, initially at rest in FC, move approximately as%
\begin{eqnarray}\label{eq:ABFC}
&& X_{A,B}=x_{A,B} - \frac{1}{2}h_{+}x_{A,B}+\frac{1}{2}h_{\times}y_{A,B}\nonumber \\
&& Y_{A,B}=y_{A,B} + \frac{1}{2}h_{+}y_{A,B}+\frac{1}{2}h_{\times}x_{A,B}\\
&& Z_{A,B}= z_{A,B}\nonumber
\end{eqnarray}
In this case, the proper distance does not change while the distance between
the two particles does: the index of refraction of the vacuum is 1. Time of
flight of a photon between two test masses $A$ and $B$ is the same in both
systems: it is a physical result, indeed, in spite of the different
descriptions.

It is to be noted that the particle $A$ moves with respect to the particle $B$
as subjected to the $"$tidal$"$ force $F_{A,i}=\frac{1}{2}m_A
x^j\frac{d^2h_{ji}}{dt^2}$.

From Eq.~(\ref{eq:ECFC}), it appears that in the study of interaction of GW
with experimental devices, Einstein coordinates are not the most suitable
because of the very complex motion resulting for an observer in the laboratory
frame; Fermi coordinates can indeed be considered as the natural extension of a
"Cartesian inertial system" of the local observer
\cite{Gualdi1982,Flores1986,Faraoni1992}.

\section*{Appendix B: Gauge invariance}\label{par:gauge}

Demonstrating the invariance of the results in Sect.~3 under general gauge
transformations is the subject of ongoing work; here we restrict the discussion
to the so called "long wavelength approximation" \cite{Flores1986,Faraoni1992}.

It is easy to see that the transformation matrix S from EC to FC of Appendix A
behaves as
\begin{eqnarray}\label{eq:S}
S=I+ O(h)
\end{eqnarray}
where I is the identity matrix; furthermore from Eq.\ref{eq:ECFC}, we get
\begin{eqnarray}\label{eq:T}
T=t+ O(hL^2/c \lambda)
\end{eqnarray}
From Eq.\ref{eq:S}, using Eq.\ref{eq:T} in the approximation $L/\lambda
\rightarrow 0$ after the insertion of the $\delta k$ term for proper laser
phases, the identity of results in both coordinates systems used in Sect.3
follows.

 It is worth noting that, starting from general FC, it is possible to build a simpler "laboratory
frame", that is a \textit{rigid coordinates system}
\cite{Misner2006,Grishchuk}, which preserves the FC properties in the
hypothesis of constant $z$ (wavefront of the gravitational plane wave) near the
$Z=0$ plane (the plane of the interferometer)
\cite{Delva2006,Grishchuk,Grishchuk2004}.
 Considering for simplicity only the $+$
polarization, the transformation law from the EC with metric
\begin{equation}\label{eq:metric2}
ds^{2}=  c^{2}dt^{2} - (1-h_{+}) dx^{2} - (1+h_{+})dy^{2} - dz^{2}
\end{equation}%
to the rigid system, is
\begin{eqnarray}\label{eq:ECFC2}
&& t=T \nonumber\\
&&x=X + \frac{1}{2}h_{+}X \nonumber \\
&&y=Y - \frac{1}{2}h_{+}Y \\
&&z=Z \nonumber
\end{eqnarray}
with transformation matrix
\begin{eqnarray}\label{eq:ABCDs}
 \left( \begin{array}{cccc}
 1 & 0 & 0 & 0 \\
 \frac{1}{2c}\dot{h}_+ X & 1+\frac{1}{2}h_+ & 0 & -\frac{1}{2c}\dot{h}_+ X \\
  -\frac{1}{2c}\dot{h}_+ Y & 0 & 1-\frac{1}{2}h_+ & \frac{1}{2c}\dot{h}_+ Y\\
  0 & 0 & 0 & 1
\end{array}\right)
\end{eqnarray}%
and metric
\begin{eqnarray}\label{eq:metrics}
&&ds^{2}=c^{2}dT^{2} - dX^{2} - dY^{2} - dZ^{2} + \\
&&+ \frac{\dot{h}_+}{c}\left(XdXdZ - YdYdZ - cXdXdT + c Y dY dT
\right)\nonumber
\end{eqnarray}%
Writing proper ABCD matrices for the rigid coordinates system by using
 Eq.~\ref{eq:metrics}, the same
results as in Eq.~\ref{eq:DRBFCOm} or Eq.~\ref{eq:DDIFCOm} can be obtained,
thus demonstrating the identity of the results in rigid, Fermi, and Einstein
coordinates systems.

\clearpage

\end{document}